# Measurement of the Flux of Ultra High Energy Cosmic Rays by the Stereo Technique

R.U. Abbasi[a], T. Abu-Zayyad[a], M. Al-Seady[a], M. Allen[a], J.F. Amann[b], G. Archbold[a], K. Belov[a,d], J.W. Belz[a], D.R. Bergman[d], S.A. Blake[a], O.A. Brusova[a], G.W. Burt[a], C. Cannon[a], Z. Cao[a,h], W. Deng[a], Y. Fedorova[a], J. Findlay[a], C.B. Finley[g], R.C. Gray[a], W.F. Hanlon[a], C.M. Hoffman[b], M.H. Holzscheiter[b], G. Hughes[d], P.Hüntemeyer[a], D. Ivanov[d], B.F. Jones[a], C.C.H. Jui[a], K. Kim[a], M.A. Kirn[c], E.C. Loh[a], M.M. Maestas[a], N. Manago[e], L.J. Marek[b], K. Martens[a], J.A.J. Matthews[f], J.N. Matthews[a], S.A. Moore[a], A. O'Neill[g],C.A. Painter[b], L. Perera[d], K. Reil[a], R. Riehle[a], M.D. Roberts[a], D. Rodriguez[a], M. Sasaki[e], S.R.Schnetzer[d], L.M. Scott[d], G. Sinnis[b], J.D. Smith[a], R. Snow[a], P. Sokolsky[a], R.W. Springer[a], B.T. Stokes[a,d], S.R. Stratton[d], J.R. Thomas[a], S.B. Thomas[a], G.B. Thomson[d], D. Tupa[b], L.R. Wiencke[a], A. Zech[d], B.K. Zhang[i], X. Zhang[e], Y. Zhang[i]

The High Resolution Fly's Eye Collaboration

[a]University of Utah,
Department of Physics and High Energy Astrophysics Institute,
Salt Lake City, Utah, USA
[b]Los Alamos National Laboratory,Los Alamos, NM, USA
[c]Montana State University, Department of Physics,
Bozeman, Montana, USA
[d]Rutgers-The State University of New Jersey,
Department of Physics and Astronomy,
Piscataway,New Jersey, USA
[e]University of Tokyo,
Institute for Cosmic Ray Research,
Kashiwa, Japan
[f]University of New Mexico,Department of Physics and Astronomy,
Albuquerque,New Mexico, USA
[g]Columbia University,Department of Physics and Nevis Laboratory,
New York,New York, USA
[h]Institute of High Energy Physics, Beijing, China

**Abstract**  The High Resolution Fly's Eye (HiRes) experiment has measured the flux of ultrahigh energy cosmic rays using the stereoscopic air fluorescence technique.  The HiRes experiment consists of two detectors that observe cosmic ray showers via the fluorescence light they emit.  HiRes data can be analyzed in monocular mode, where each detector is treated separately, or in stereoscopic mode where they are considered together. Using the monocular mode the HiRes collaboration measured the cosmic ray spectrum and made the first observation of the Greisen-Zatsepin-Kuzmin cutoff. In this paper we present the cosmic ray spectrum measured by the stereoscopic technique. Good agreement is found with the monocular spectrum in all details.

## I.   Introduction

The spectrum of ultra high energy (UHE) cosmic rays is a rich field for study. It is characterized by several features (see Fig. 9) including the second knee at about $3 \times 10^{17}$ eV, the ankle at $4.5 \times 10^{18}$ eV, and the Greisen-Zatsepin-Kuzmin (GZK) cutoff at $5.6 \times 10^{19}$ eV. The last feature was predicted by K. Greisen [1], and G. Zatsepin and V. Kuzmin [2], and is caused by the energy threshold for pi meson production in interactions between cosmic ray protons of extragalactic origin and photons of the cosmic microwave background radiation (CMBR). This is a strong energy loss mechanism for protons, and limits their range to about 50 Mpc from Earth. The ankle is likely caused by electron-positron pair production in these same interactions [3]. This process is a weaker energy loss mechanism and the horizon of cosmic ray protons in the ankle region is much larger. Some controversy exists about this interpretation of the ankle, with another possible mechanism being the transition between cosmic rays of galactic origin to an extragalactic flux. The energy of the second knee has large uncertainties, and its cause is completely unknown. The elucidation of the details of these features, together with an investigation of composition of cosmic rays as a function of energy and possible anisotropies, forms the focus of all experiments in this energy regime.

## II.   The HiRes Experiment

The High Resolution Fly's Eye (HiRes) experiment has been previously described [4] [5]. HiRes consists of two detectors detecting UHE cosmic ray showers by collecting the emitted fluorescence light. The detectors are located atop two hills 12.6 km apart in the west-central Utah desert, and were operated on clear, moonless nights over a period of nine years (1997-2006).

The first detector deployed, called HiRes-I, consists of 21 telescopes together covering a field of view from 3° to 17° in elevation and 336° in azimuth. Each telescope is composed of a spherical mirror, of effective area 3.8 $m^2$ which focuses the showers' fluorescence light on a camera which is a cluster of 256 photomultiplier tubes. The effective area takes into account shadowing by the camera. The tubes are hexagonal in cross section so they can be close packed, and each subtends 1° by 1°. Each camera thus subtends a 16° by 16° section of the sky. As a cosmic ray shower proceeds downward through the atmosphere, its image in fluorescence light moves up across the cluster of phototubes. Timing and pulse height information from the phototubes is saved for later analysis by a sample-and-hold electronics system. HiRes-I was operated for the full nine years of the experiment's life.

The second detector, called HiRes-II, consists of 42 telescopes that cover 3° to 31° in elevation and 352° in azimuth. A flash ADC readout-system operating at 10 MHz is used to save the timing and pulse height information. HiRes-II operated from December, 1999, until April 2006. In the interval from Dec, 1999 to April 2006 relevant for the stereo spectrum, HiRes-I took data for 4522 hours and HiRes II for 4064 hours. Coincident stereo operation time was 3460 hours.

We analyze the HiRes data in two ways. In monocular mode we use the information from each detector independently. Previous HiRes spectrum publications [11] [12] reported monocular mode results. This method yields two spectra that together have the best statistical power and widest energy range. Because the HiRes1 detector ran for the longest period of time its data have the best statistics at the highest energies. However its limited elevation coverage means that events below about $10^{18.5}$ eV tend to have shower maxima outside the field of view, making their energies difficult to reconstruct. HiRes2, on the other hand, covers higher elevation angles and can reconstruct events down to $10^{17.2}$ eV.

Using the information from the two detectors simultaneously allows the reconstruction of cosmic ray showers in stereoscopic mode. HiRes stereo data have the best energy resolution. In addition, stereo reconstruction provides an important check on our understanding of the detectors response and event reconstruction. One can make two determinations of the energy, and depth of maximum shower development, and thus can measure the resolution in these two quantities entirely from the experimental data. In this paper we briefly summarize the previously published monocular reconstruction, present the energy spectrum measured using the stereo technique and compare it to the monocular result.

## III. HiRes Monocular Reconstruction

In monocular reconstruction, one starts by calibrating the data (see the next section), followed by pattern recognition. Tubes are determined to be part of the track if they are contiguous with their neighbors in the angles they subtend, and in time. Next one determines the geometry of the event (direction of the shower axis in space, and distance from the detector). The plane that contains the shower and the detector is found from a fit to the tube pointing directions. The angle of the shower in that plane (called psi) and the shower's impact parameter in the plane are found from a fit to the phototube signal relative times versus their pointing angles in the shower-detector plane. For the HiRes-II detector the resolution in psi is about 5°. For HiRes-I, the track length is typically shorter than those seen by HiRes-II which makes the time fit less unique, so a constraint is added to the fit to phototube times and angles. This constraint compares the shape of the shower development profile to the Gaisser-Hillas formula [6]. This formula has been shown to fit measured shower profiles and profiles of simulated showers well, with small fluctuations about the mean [7] . The result of this "profile constrained" fit is a resolution in psi of about 7°.

With the shower geometry known, one can calculate the shower development profile. The integral of this profile gives the calorimetric energy deposited by the shower in the atmosphere. Corrections for neutrinos and muons (so-called "missing energy") are made as indicated in the next section. The resulting resolution in energy is typically about 15%.

To calculate the aperture of the experiment we perform a complete simulation of the response of the detector to cosmic ray showers. The standard Corsika [8] program is used to generate simulated proton and Fe showers in the atmosphere and the generation and transmission of fluorescence and Cherenkov light is calculated followed by a complete simulation of the detector and electronics response. This includes wavelength dependent calculation of air fluorescence and Cherenkov light production, transmission of light through the atmosphere taking into account molecular and aerosol scattering, collection of light by the mirrors using ray tracing and the response of the photomultipliers and electronics. Simulated events are recorded in the same format as real data and processed in an identical fashion.

The previously measured cosmic ray spectrum [9] is used to weigh the energy distribution of simulated showers. We assume a proton dominated composition since this is in good agreement with both the mean value of Xmax and its measured fluctuations [10] (see Fig. 4). The Monte Carlo event distributions in all variables look just like those of the data. As an example, Figure 1 shows a comparison of the data and Monte Carlo histograms of the zenith angle of events seen by the HiRes-I detector. The excellent agreement between the Monte Carlo and the data is characteristic of our analysis as a whole.

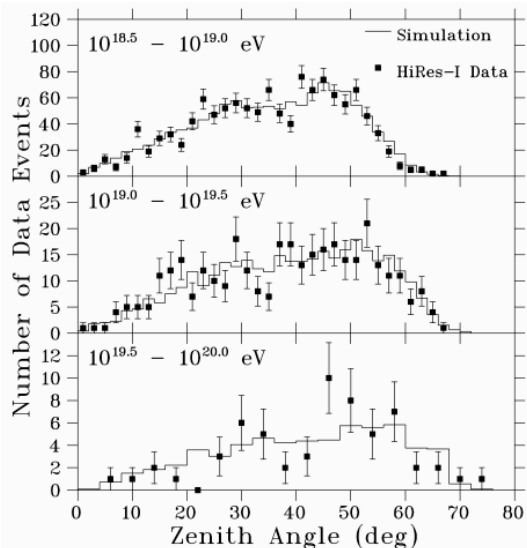

**Figure 1. Comparison of data and Monte Carlo histograms of the zenith angle of cosmic ray showers as seen by the HiRes-I detector. Comparisons in three energy ranges are shown.**

## IV. Corrections and Calibrations

The photometric calibration of the HiRes telescopes has been described previously [13]. It is based on a very stable xenon flash lamp that is placed at the center of the mirror and illuminates the phototubes. This lamp is carried from mirror to mirror on a monthly basis. The absolute brightness of the lamp has been measured using a NIST-calibrated hybrid photodiode. A detailed description of the calibration can be found in [18]. The photonic scale of the HiRes phototubes has been measured using the absolute brightness of the xenon flash lamp, and by using photon statistics, and the two results are consistent. Relative nightly calibrations were performed using a yttrium aluminum garnet (YAG) laser whose pulses were distributed to the phototube clusters via a quartz fiber system. In addition, the overall end to end optical calibration was checked by reconstructing scattered light from a pulsed N2 laser placed ~3.5 km from the two detector sites and fired into the atmosphere. Using these methods we achieve ~10% rms accuracy in our photometric scale.

Both the molecular and aerosol components of the atmosphere scatter the fluorescence light and a correction must be made for this effect. The Raleigh scattering cross section which describes the molecular scattering is well known, and by using measurements of atmospheric density made by the U.S. Weather Service at airports near the HiRes site, we can make an accurate estimate of scattering due to this effect. The aerosol component of the atmosphere can be very variable. We measure the scattering properties of aerosols at the HiRes site on an hourly basis. Steerable lasers fire patterns of shots that cover the HiRes aperture, and the HiRes detectors themselves measure the intensity of the scattered light. The most important parameter we measure is the vertical aerosol optical depth (VAOD). The mean VAOD is 0.04 with an rms variation of 0.02 over most of the lifetime of the HiRes experiment. The atmosphere at our site [14] is much clearer than the "standard desert atmosphere" used by meteorologists, which has a mean VAOD of 0.10. The VAOD at the HiRes sites is very stable, typically with only small variations in VAOD over one or several nights. Since the first 2.5 years of HiRes-I data were collected before the atmospheric measurement system was deployed, we used the average VAOD instead of the hourly measurements in the monocular analysis reported previously. Studies show that replacing this by the hourly aerosol database has a negligible affect on the monocular cosmic ray spectrum [15]. We use the same technique in the stereo analysis reported here for consistency.

The presence of clouds can affect the energy determination of an event and the detector aperture. The HiRes experiment deployed sensitive infrared detectors to monitor changes in the sky temperature produced by clouds as a function of time. To minimize possible systematic effects in the stereo analysis we require cloud free conditions in our final data cuts, as determined by the infrared monitors and by visual observations.

The intensity of fluorescence light emitted from a cosmic ray shower is proportional to the total ionization energy deposited by the charged particles in the shower [16]. The proportionality constant is called the fluorescence yield, and has been measured by several groups [17]. For this analysis we used the average of the measurements from the papers of reference [16] for the total fluorescence yield and use a spectral distribution of light given by the Bunner [18]. This average is essentially identical to the yield reported by Kakimoto et al. in [17] and is hence consistent with previously published monocular HiRes analyses which used this value. The error on the mean, based on the individual quoted errors, is 6%. We use this as a measure of the systematic uncertainty on this yield.

By using the average dE/dx determined from Corsika shower simulations we can calculate the number of charged particles in the shower as a function of slant depth, and hence the calorimetric energy of the shower. We also calculate the average "missing energy" i.e., that which goes into neutrinos and muons and is not included in the calorimetric energy, from simulations. We find that it is about 10% [15]. We apply this correction to the calorimetric energy to obtain our estimate of the energy of the primary cosmic ray. We include the uncertainties in fluorescence yield, mean dE/dx, and missing energy in our budget of systematic uncertainties in energy.

### V.    Stereo Event Reconstruction

For the stereo analysis the shower geometry is determined primarily by finding the intersection of the two shower-detector planes. The shower axis is located at this intersection. Phototube timing plays a minor role here. The shower-detector plane from each site is determined by a fit to the weighted direction vectors of the pixels. The fit is improved slightly by using the HiRes-II phototube timing. This process results in a mean resolution in shower arrival direction of about ½ degree.

Because the tracks seen by the HiRes-II detector are typically longer than those from HiRes-I, the shower profile is measured from the HiRes-II data. Here the pulse height signals in all tubes in each FADC time bin are added, and the list of pulse heights as a function of time forms the basis of an iterative calculation of the shower's calorimetric energy and $X_{max}$. In this calculation values of energy and $X_{max}$ are assumed, and, using the event geometry, atmospheric corrections, and photon calibration, the list of pulse heights is predicted. A shower profile shape of a Gaussian in age is assumed.

Upon comparison with the data, the values of energy and $X_{max}$ are corrected and the calculation repeated until it converges. The correction for missing energy is then made.

The HiRes-I detector signals are treated in an identical way except that each phototube signal is treated individually. Although the HiRes-I profiles are not used in reconstructing cosmic ray energies, there exists a subset of events where the HiRes-I energy measurement is excellent. This subset is used to measure the resolution in energy and $X_{max}$ experimentally.

Minimal cuts on the raw data to begin analysis include: successful fit to tube directions to determine the shower-detector plane with a minimum angular track length of six degrees and track tube timing consistent with downward pointing track. Stereo events are assembled by requiring an event time coincidence as described below and further analysis cuts are described in section VI.

Figure 2 shows the profiles of a typical stereo event. The profiles from both HiRes-I and HiRes-II are shown.

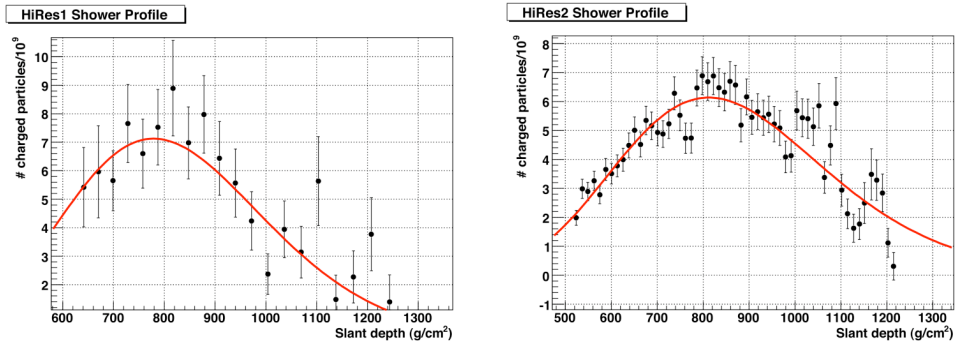

**Figure 2. Typical shower profile for stereoscopic reconstruction. Profiles seen by the HiRes-I and HiRes-II detectors are shown. This event has an impact parameter of 18.6km from HiRes-I and 14.4 km from HiRes-II and a zenith angle of 53.6 degrees. It has a reconstructed energy of $8.6 \times 10^{18}$ eV from HiRes-I and $8.4 \times 10^{18}$ eV from the HiRes-II profile.**

Observing events stereoscopically allows us to measure the energy resolution by comparing energies measured by HR1 and HR2 independently for the same event. The distribution in energy differences for real and simulated data is shown in Fig. 3. Here, in addition to the standard analysis cuts described in VI, we require that both HiRes-1 and HiRes-2 see the shower maximum. This distribution represents the convolution of the energy resolutions of the individual detectors. The width of the data distribution of 33% is consistent with the predicted width from simulated showers of 30%. This implies that the detector simulation adequately represents the actual detector resolution.

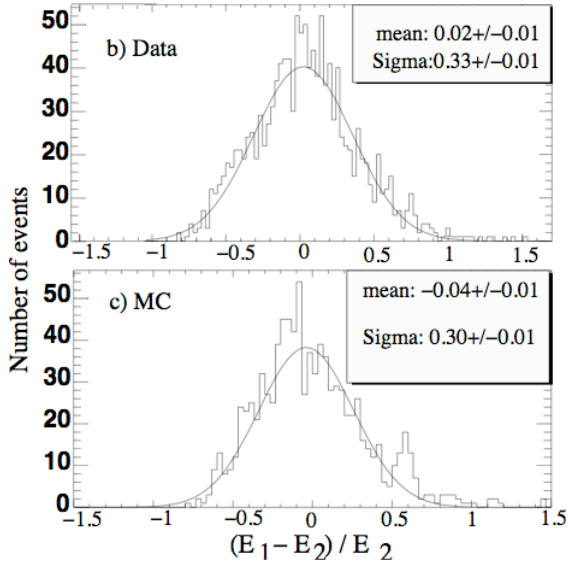

Fig. 3 Fractional difference between HiRes-1 and HiRes-2 measurements of individual shower energies. Top: data; bottom: Monte Carlo (MC) simulation.

## VI Stereo Detector Aperture Calculation

The detector aperture is defined as the effective area times solid angle in which an air shower of a given energy will trigger the detector, and survive the quality cuts to be included in the final event sample described in VI. For each of the HiRes detectors, the aperture grows with energy since higher energy showers are brighter and can be detected at larger distances. The apertures saturate at the highest energies to a value approaching 10,000 $km^2 sr$ mostly due to the effect of decreased angular extent of the showers at large distances from the detectors. The shower detection efficiency for each detector is calculated using a detailed detector simulation program driven by proton showers from the Corsika program. All experimental conditions are simulated in the Monte Carlo program just as they are found in the experiment. These conditions are identical to those used in the monocular analysis. The detector triggering is realistically modeled and events are reconstructed using exactly the same programs as for real showers.

A detailed comparison between simulated and measured event distributions is carried out to ensure that the simulation reflects reality as much as possible. An example, similar to many, is shown in Fig. 4. The stereo aperture is determined by demanding that both HiRes-I and HiRes-II detectors trigger on the same event.

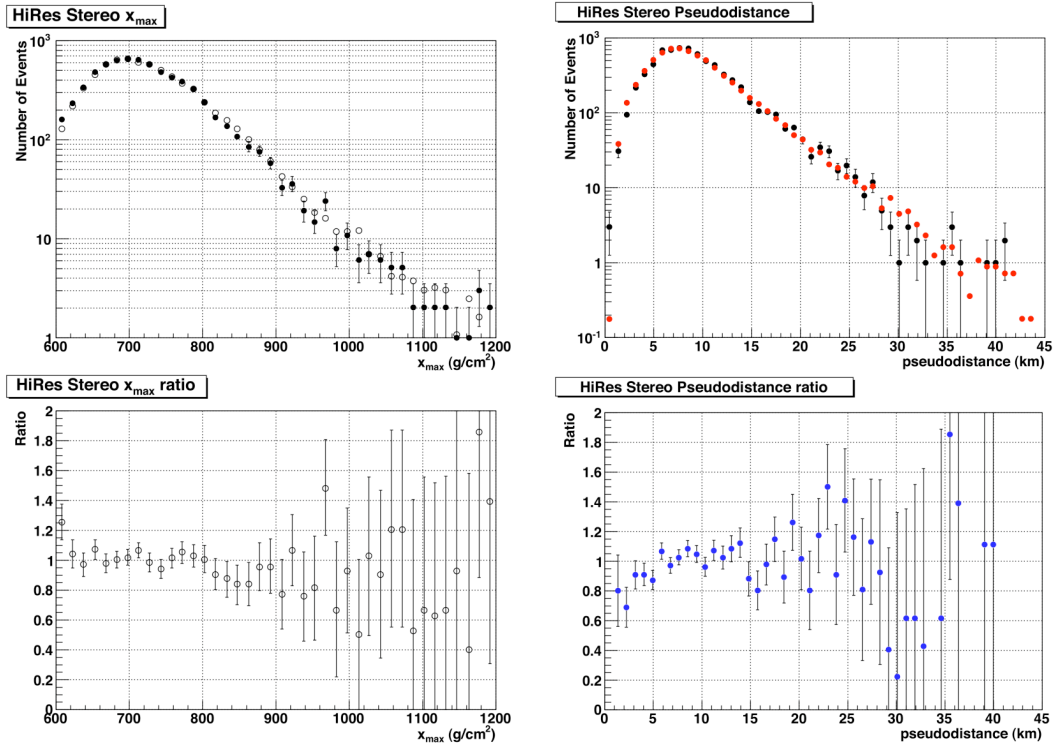

**Figure 4.** Comparison of Monte Carlo and data histograms of the Xmax distribution and the impact parameter (from HiRes-II) of events, showing excellent agreement, which is characteristic of our Monte Carlo simulation as a whole. The input for the Monte Carlo assumes a purely protonic composition.

The resulting aperture is shown in Fig. 5. The stereo aperture has a similar form to the monocular apertures but drops significantly below the monocular apertures for energies below $10^{18.5}$ eV. Events with energies below this value can only trigger both detectors if they appear in a volume midway between the two detectors. This volume rapidly goes to zero at lower energies, while the individual detector apertures remain finite.

The selection criteria applied to the HiRes-I and HiRes-II monocular analyses are more restrictive than that of the stereo analysis because of the nature of the timing fits employed. This has the result that the stereo aperture is a bit higher than the mono apertures above about $10^{19.7}$ eV.

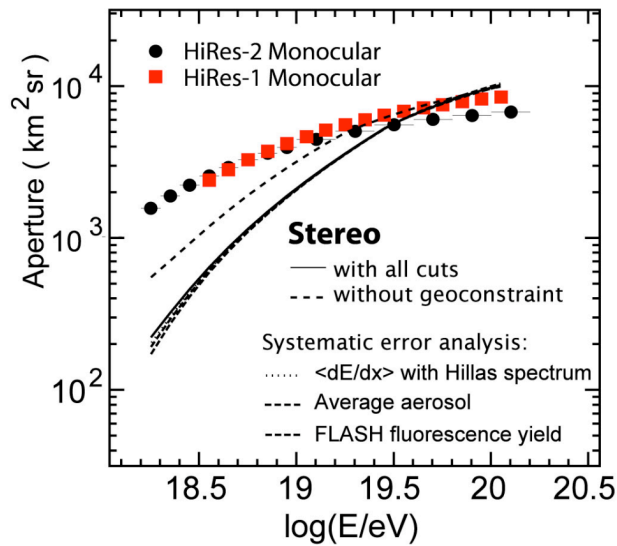

Fig. 5 HiRes stereo apertures for different assumptions. The dashed line represents the regular aperture and the other lines in the figure show the aperture when one applies the geometrical constraint described in the text. Varying the mean dE/dx of showers, the average aerosol level, or the fluorescence yield, has a minimal affect on the aperture. The apertures of the HiRes-1 and HiRes-2 detectors, operating in monocular mode, are also shown. Geoconstraints are described in the text below.

Our calculation of the aperture can be checked by imposing additional constraints. Simulations show that within a constant distance between our detectors and showers we collect cosmic ray events with 100% efficiency above a certain energy. For instance if Rp , defined in this case as the impact parameter of the shower from the centroid between the two detectors, is less than 10 km, all events above $10^{18.2}$ eV are collected with nearly 100% efficiency, and for 20 km the corresponding energy is $10^{19.0}$ eV. This information can be read from Figure 6. This defines a "geoconstraint" which allows us to calculate the aperture (and spectrum) at all energies with 100% efficiency.

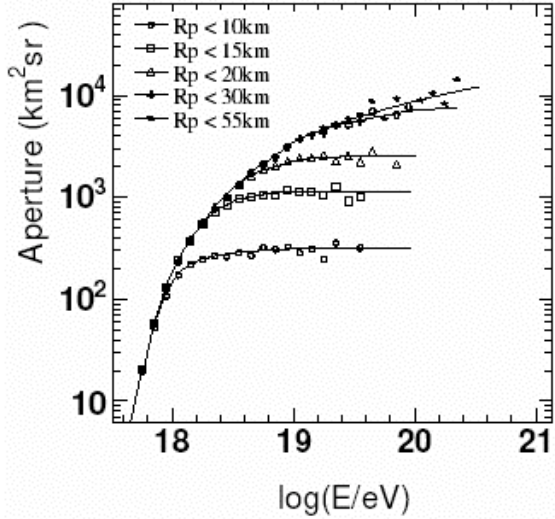

Fig. 6 HiRes stereo aperture for events restricted to less than various Rp values.

## VI.  Spectrum and Discussion

Events that passed minimum analysis criteria for both detectors were time matched, individual event-detector planes were determined for each detector and the event geometry and shower profiles reconstructed.  The stereo aperture calculation becomes robust at energies greater than $10^{18}$ eV, so we impose an overall energy cut on the data of $E > 10^{18.2}$ eV. The following cuts were applied to the reconstructed events: cloud free observation time (69% of events passed this cut), successful profile fit with Xmax in field of view of HiRes-II (79%), Cherenkov contamination less than 30% of total photons (93%).  These cuts ensure that the detector aperture used in the spectrum determination is not affected by unknown cloud distributions, that the shower profile is unaffected by clouds, that it is well determined near the shower maximum, and that biases on the shower profile shape due to large subtractions of Cherenkov light are minimized.  2267 events survived these cuts. While the minimum track length for analysis is six degrees, the requirement for a successful shower profile fit with Xmax in the HiRes-2 field of view results in an angular track length distribution ranging from 10 to 50 degrees with a mean of 24.7 deg. This track length distribution agrees well with predictions from  simulated events. The aperture and resulting spectrum with these cuts are shown in Figs. 5 and 7.

The most robust stereo spectrum results from the additional application of the geometric constraint described above.  The addition of this constraint reduces the data sample to 1147 events. This "fully efficient" spectrum is shown in Fig. 9. As discussed previously, this spectrum is insensitive to variations in atmospheric attenuation.

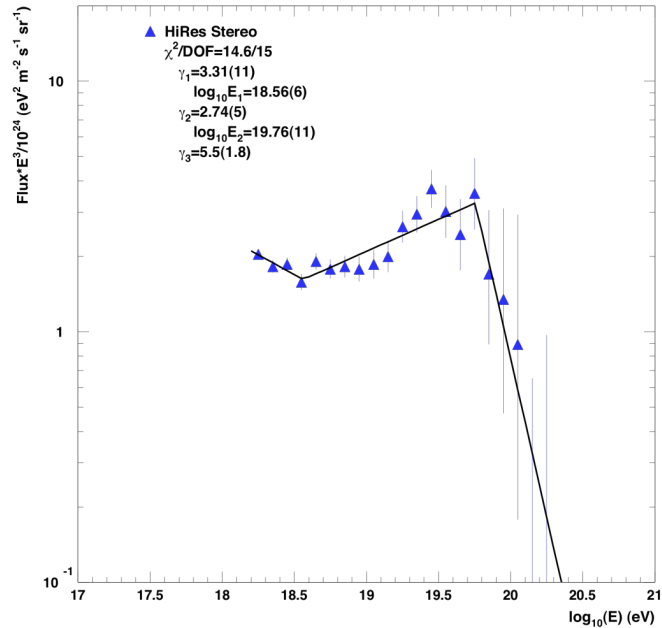

**Figure 7. HiRes stereo spectrum and power law fits. The lines are a fit described in the text.**

Monte Carlo studies using events from a Corsika simulation of air showers, using the QGSJet01 hadronic generator routine, indicate that there is a (~20%) sensitivity to the primary composition (protons or iron) below about $10^{18.4}$ eV through the composition dependence of the aperture calculation. Above $10^{19}$ eV the spectrum is certainly independent of primary composition. However, since the measured Xmax distribution is well described by pure protons with an upper bound on the iron fraction of 10% down to $10^{18.2}$ [20], this systematic error can be conservatively reduced to 5% over the entire energy range.

Previous HiRes publications on the cosmic ray spectrum reported the observation of the ankle [10], and announced the first observation of the GZK cutoff [12]. The most recently published HiRes monocular spectra are displayed in Figure 8. These spectra correspond to on-times of 4002 and 1397 hours for HiRes-I and HiRes-II respectively. A good fit to the data can be made using three power-law segments, shown as black lines on the figure. This fit has break points at log(E) (E in eV) of 18.65 ± 0.05 (the energy of the ankle) and 19.75 ± 0.04 (the energy of the GZK cutoff). The three power laws are -3.25 ± 0.01, -2.81 ± 0.03, and -5.1 ± 0.7, respectively below the ankle, between the ankle and the cutoff, and above the cutoff.

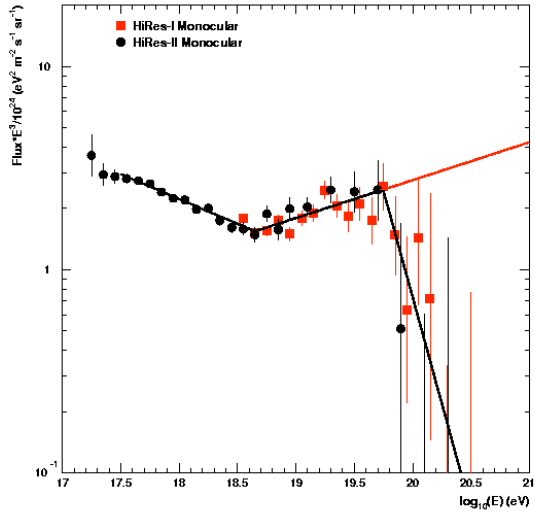

**Figure 8. Cosmic Ray Spectrum Measured by the HiRes Experiment in Monocular Mode. The spectrum from the HiRes-I (red) and HiRes-II (black) are shown. $E^3$ times the flux vs. the log of the energy is plotted vs. the log of the energy. The black lines are a fit to the data that is described in the text.**

The spectrum in Fig 7, while having a much higher minimum energy cut, shows an ankle structure that is consistent with the monocular HiRes-I and HiRes-II spectra shown in Fig. 8. We performed a fit identical in all details to the monocular fits previously reported to the ankle region. This fit finds the ankle at $10^{18.56}$ eV. The two power law indexes are found to be –3.25 ± .11 and -2.81 ± .05 below and above the minimum respectively. The significance of the ankle in this fit is 4.8 σ. The geometrically constrained spectrum in Fig. 8 also shows a similar pronounced ankle structure.

The GZK suppression that is seen in the monocular spectrum ( Fig. 8 ) is also apparent in the stereo spectrum, which shows a deficit of events above $10^{19.8}$ eV. To test the statistical significance of this suppression we extend the broken power law fit beyond this energy and calculate the expected number of events above $10^{19.8}$ if the spectrum were to continue in this manner. For the most reliable geo-constrained spectrum we find that 27 events would be expected as compared to 7 events measured. This corresponds to a 3.8 sigma effect.

The main systematic effects related to the stereo energy reconstruction are the uncertainties in the photonic scale (10%), fluorescence yield (6%), the calculation of deposited energy as a function of atmospheric depth which depends on the value of the mean dE/dX as a function of shower development depth (10%), and the aerosol concentration (6%). Added in quadrature this comes to a 17% systematic uncertainty in our energy scale. The uncertainty in the flux is 30%.

## VII. Comparison of Mono and Stereo Results

The spectra measured using the monocular and stereoscopic methods agree very well, as can be seen from Figure 9. Both analysis methods use the same input parameters (fluorescence yield, VAOD, etc.). The observed energies of the ankle and the GZK breaks agree within uncertainties, as do the power laws below the ankle, and between the ankle and the GZK cutoff. The exact values of fit values are displayed in Table 1.

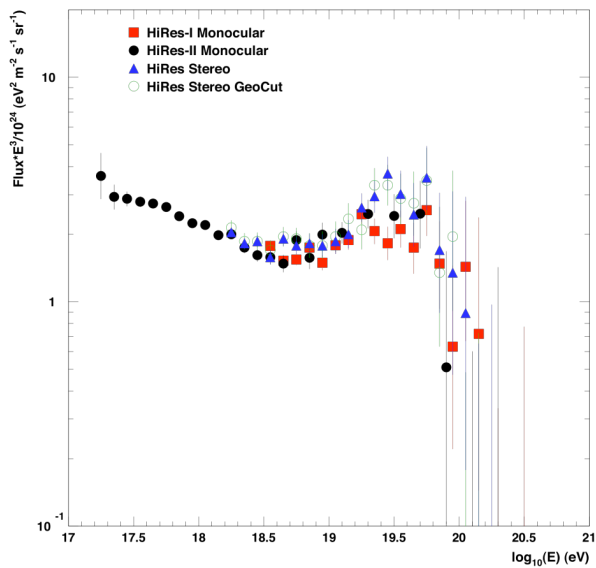

**Figure 9. HiRes monocular and stereo spectra plotted together showing the excellent agreement between the two, and the ankle and GZK cutoff features of the spectrum. The GeoCut spectrum corresponds to the geometrically constrained aperture cuts discussed in the preceeding.**

|  | Mono Spectra | Stereo Spectrum |
|---|---|---|
| Power law (below ankle) | -3.25 ± 0.01 | -3.31 ± 0.11 |
| Power law (intermediate) | -2.81 ± 0.03 | -2.74 ± 0.05 |
| Power law (above GZK) | -5.1 ± 0.7 | -5.5 ± 1.8 |
| log(Energy) of Ankle | 18.65 ± 0.05 | 18.56 ± 0.06 |
| log(Energy) of GZK break | 19.75 ± 0.04 | 19.76 ± 0.11 |

**Table 1. Results of power-law fits to the monocular and stereo spectra.**

Fig. 10 shows the HiRes stereo and mono spectra in comparison to the AGASA and Pierre Auger Observatory (PAO) [21] spectra. While 10-20% energy scale shifts can bring the bulk of these spectra into good agreement, neither the HiRes nor the PAO spectra confirms the AGASA claim of a continuing spectrum beyond the GZK cutoff.

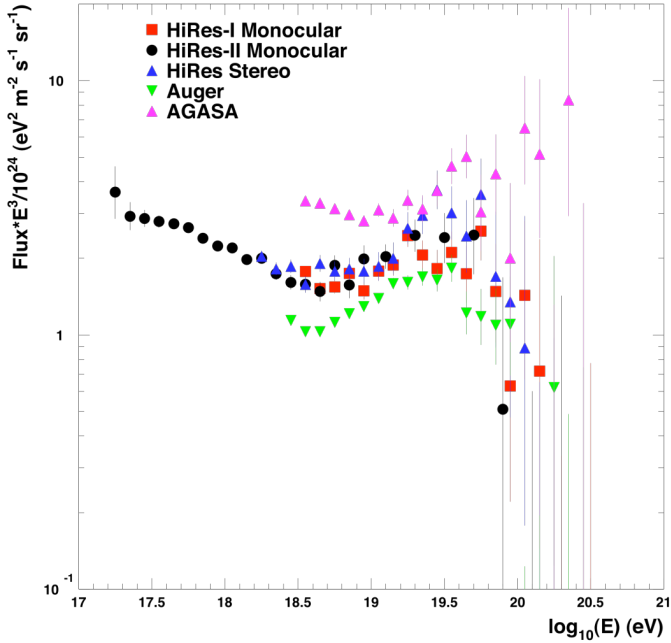

**Figure 10. HiRes monocular and stereo spectra plotted together with the AGASA and Pierre Auger Observatory (PAO) spectra.**

# VIII. Conclusions

We have measured the spectrum of ultrahigh energy cosmic rays using the stereoscopic fluorescence technique. The HiRes experiment deployed two fluorescence detectors located 12.6 km apart on the U.S. Army Dugway Proving Ground in Utah, and collected data for nine years, with both detectors running from 1999 to 2006.

Data from events seen by both of the HiRes detectors were analyzed simultaneously to reconstruct cosmic ray showers' parameters, and the calorimetric deposition of showers' energy was used to measure the energy of the primary cosmic rays. The energy resolution of stereo reconstruction is about 10%, whereas monocular reconstruction achieves about 15%. The difference comes from improved geometrical accuracy of the stereo technique. Monocular reconstruction, however, yields spectra of the best statistical power, and widest energy range. Mono and stereo are sensitive to the various sources of systematic errors in different ways, and it is very important to use both techniques to learn the most about the cosmic ray spectrum.

The spectrum calculated using the stereo technique agrees well with the monocular spectra that the HiRes collaboration has published previously. In particular the two features of the spectrum called the ankle and the GZK cutoff appear in the stereo spectrum just as they do in the monocular spectra.

## IX. Acknowledgements

This work is supported by U.S. NSF Grants Number PHY-9321949, PHY-9322298, PHY-9904048, PHY-9974537, PHY-0098826, PHY-0140688, PHY-0245428, PHY-0305516, PHY-0307098, and by the DOE Grant Number FG0392ER40732.  We gratefully acknowledge the contributions from the technical staffs of our home institutions.  The cooperation of Colonel E. Fischer, Colonel G. Harter, and Colonel G. Olsen, the U.S. Army, and the Dugway Proving Ground staff is greatly appreciated.